\documentclass{sig-alternate}
\usepackage{alltt                                    
          , multirow
          , booktabs
          , listings
          , graphicx
          ,float
	,cite
          ,verbatim
         ,mathtools
	,url
	,amsmath
}
\usepackage[table]{xcolor}
\usepackage[numbers]{natbib}     
\usepackage{syntax}
\usepackage{algorithmic, algorithm}
\usepackage{enumitem}
\usepackage{threeparttable}
\usepackage[T1]{fontenc}
\usepackage{hhline}
\usepackage{bm}
\usepackage{balance}

\usepackage{expl3}
\ExplSyntaxOn
\newcommand\latinabbrev[1]{
  \peek_meaning:NTF . {
    #1\@}%
  { \peek_catcode:NTF a {
      #1., \@ }%
    {#1., \@}}}
\ExplSyntaxOff

\newcommand{\CASE}[1]{\STATE \textbf{case} #1\textbf{:} \begin{ALC@g}}
\newcommand{\ENDCASE}{\end{ALC@g}}

\newcommand{\DEFAULT}{\STATE \textbf{default:} \begin{ALC@g}}
\newcommand{\ENDDEFAULT}{\end{ALC@g}}
\newcommand{\DEFAULTLINE}[1]{\STATE \textbf{default:} }
\let\footnotesize\scriptsize

\newsavebox{\supbox}
\newcommand{\bsup}{\begin{lrbox}{\supbox}$\tt\scriptstyle}
\newcommand{\esup}{$\end{lrbox}{}^{\usebox{\supbox}}}
\def\eg{\latinabbrev{e.g}}
\def\ie{\latinabbrev{i.e}}

\definecolor{lightpurple}{rgb}{0.8,0.8,1}
\definecolor{codebg}{RGB}{255,255,255}
\definecolor{commentcolor}{RGB}{11,140,11}
    
\newcommand{\hl}{\makebox[0pt][l]{\color{lightgray}\rule[-2.5pt]{0.98\linewidth}{9pt}}}

\usepackage{balance}

\begin{document}
%
\permission{Copyright $\textcopyright$ 2015 Mohammad Rahman and Chanchal Roy. Permission to copy is hereby granted provided the original copyright notice is reproduced in copies made.}
\copyrightetc{}


\title{Recommending Relevant Sections from a Webpage about Programming Errors and Exceptions}
%
%
%
%
%

\numberofauthors{2}
\author{
\alignauthor Mohammad Masudur Rahman\\
       \affaddr{Department of Computer Science}\\
      \affaddr{University of Saskatchewan, Canada}\\
       \email{masud.rahman@usask.ca}
\alignauthor Chanchal K. Roy\\
       \affaddr{Department of Computer Science}\\
       \affaddr{University of Saskatchewan, Canada}\\
       \email{chanchal.roy@usask.ca}
}



\maketitle
\begin{abstract}

Programming errors or exceptions are inherent in software development and maintenance, and given today's Internet era, software developers often look at web for finding working solutions. They make use of a search engine for retrieving relevant pages, and then look for the appropriate solutions by manually going through the pages one by one. However, both the manual checking of a page's content against a given exception (and its context) and then working an appropriate solution out are non-trivial tasks.  
They are even more complex and time-consuming with the bulk of irrelevant (\ie\ off-topic) and noisy (\eg\ advertisements) content in the web page. In this paper, we propose an IDE-based and context-aware page content recommendation technique 
that locates and recommends relevant sections from a given web page by exploiting the technical details, in particular, the context of an encountered exception in the IDE. An evaluation with 250 web pages related to 80 programming exceptions, comparison with the only available closely related technique, and a case study involving comparison with VSM and LSA techniques show that the proposed technique is highly promising in terms of precision, recall and $F_1$-measure. 

\end{abstract}




%


\section{Introduction}
Studies show that about 80\% of total effort is spent in software maintenance \cite{surfclipse}. During the development and maintenance of a software product, software developers deal with different programming errors and exceptions, and they often search in the web for working solutions for solving them. According to \citet{twostudy}, developers spend about 19\% of their development time in web surfing. 
During the collection of information using traditional web search, they first use a search engine with a few keywords for retrieving  relevant pages. However, in order to locate the required information, they need to go through the pages one by one, which is challenging, and this paper focuses on this particular research problem.
Both manual checking of a web page for relevant content against an error and working an appropriate solution out are non-trivial tasks. 
These tasks get even more complex and time-consuming with the bulk of irrelevant (\ie\ off-topic) and noisy (\eg\ advertisement) content in the page.
As early as 2005, \citet{gibson} estimated that about 40\%-50\% of web data were simply noise.
Thus, the developers often spend a significant amount of time and efforts in searching and then extracting the content of interest from the web pages. 
Fortunately, automated support in post-search content analysis can greatly benefit them in this regard. For example, identification and then recommendation of page sections relevant for the developers from a selected web page can help them get rid of information overload and locate the content of interest instantly, which in turn reduces their overall problem-solving efforts.

\begin{figure*}[!tb]
\centering
\includegraphics[width=6.2in]{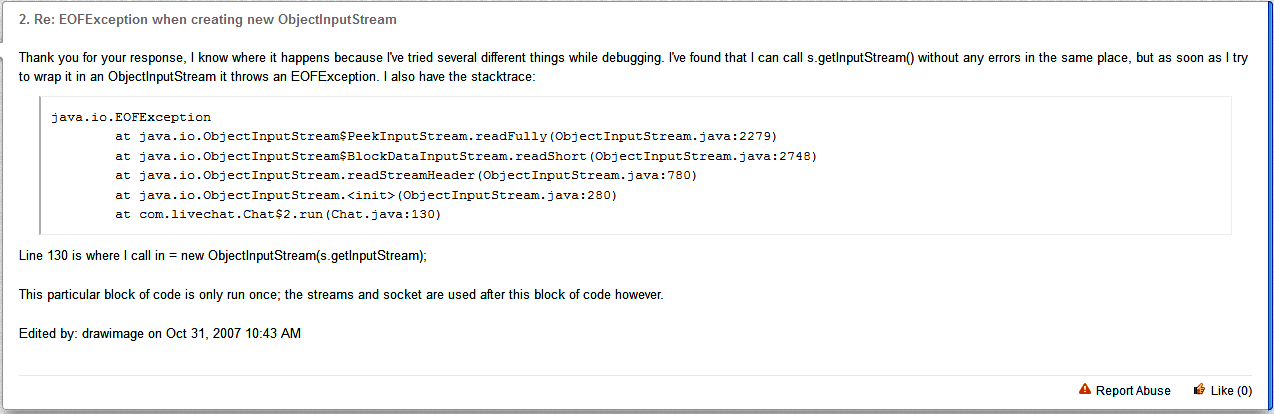}
\vspace{-.3cm}
\caption{An Example Relevant Section}
\label{fig:relsec}
\vspace{-.6cm}
\end{figure*}

A number of existing studies focus on extracting the noise-free version of a web page by applying different techniques \cite{boilerplate,cafella,mladv, mlnepo, densitometric,little,maincontent, sun, vips, dsc, ccb}.
However, no studies target the extraction of relevant sections or sections of one's  interest from the page. 
Thus, they fail to direct one to the right (or relevant) sections, and do not help much either in reducing information overload or in locating solution in the page. 
Furthermore, most of the techniques are domain specific (\ie\ applies domain knowledge) or template specific (\eg\ tabular structure) \cite{boilerplate, cafella}, and they extract content from various domains such as news \cite{maincontent, sun, dsc, ccb,densitometric}, Wikipedia \cite{cafella}, and real estates \cite{little}. However, none of them deals with programming related web pages, which makes our work unique and novel in this context.

\begin{figure}[!tb]
\centering
\includegraphics[height=4in, width=2.5in]{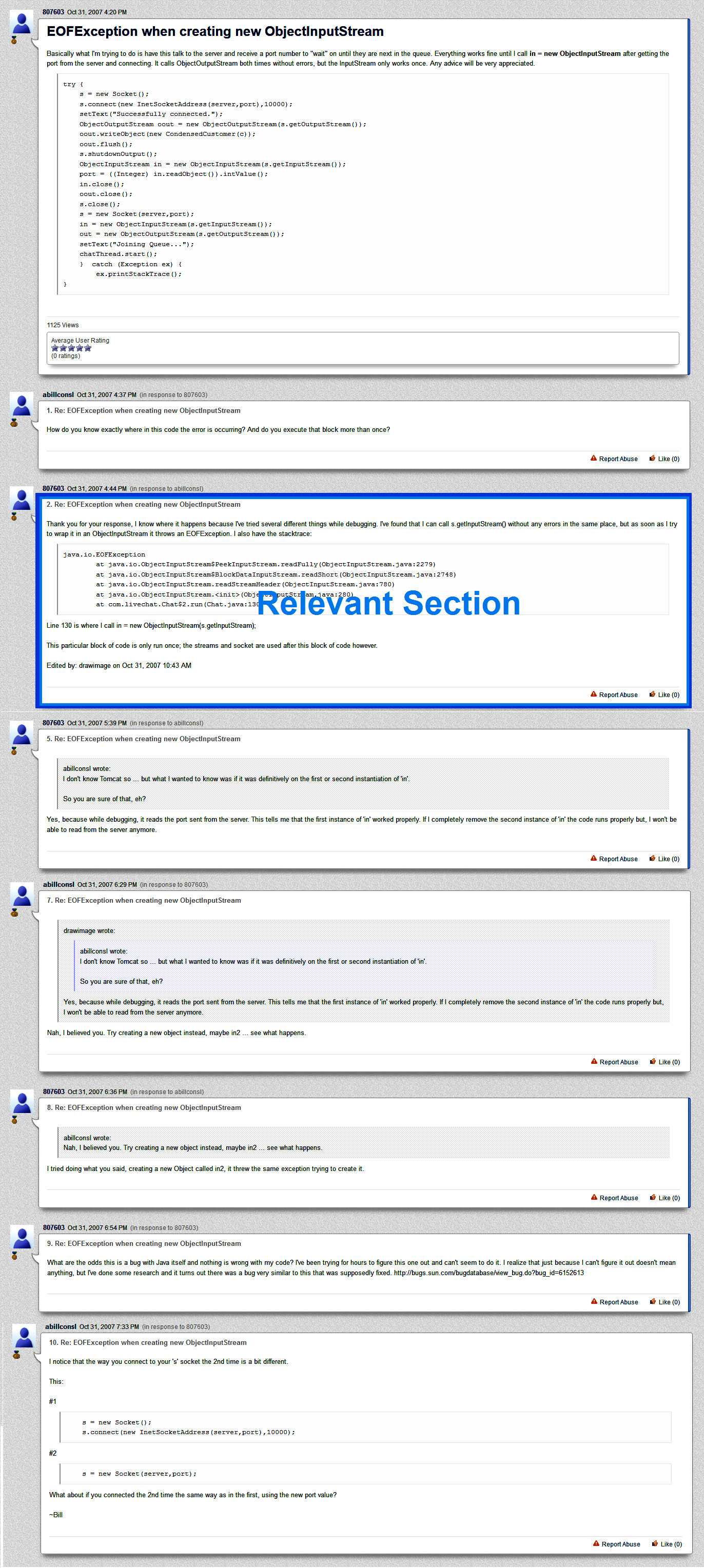}
\vspace{-.3cm}
\caption{Relevant Section(s) in the Webpage}
\label{fig:moti}
\vspace{-.5cm}
\end{figure}

In this paper, we propose a novel technique that identifies and then recommends the relevant sections from a programming related web page by exploiting the technical details of an encountered exception in the IDE (\ie\ \emph{context-aware} technique).
Once a developer searches about an exception using a few keywords, the search engine (\eg~in our case Google) returns a number of pages. Then the real challenge for her is to manually check those pages and collect meaningful information for the encountered exception, where our technique comes handy.
For example, the code under development (hereby we call it \emph{context code}) in Listing \ref{lst:ccontext} triggers an \emph{EOFException}, and the IDE reports the stack trace in Listing \ref{lst:strace}.
Our technique analyzes both the content of a returned web page (\eg\ Fig. \ref{fig:moti}) and the technical details of the exception (\eg\ Listing \ref{lst:ccontext} and Listing \ref{lst:strace}), analyzes legitimacy (\ie\ content purity) and relevance of different sections from the page, and then identifies the most relevant section (Fig. \ref{fig:relsec}, boxed area from Fig. \ref{fig:moti}) in the page for the developer.
We integrate Google search API into Eclipse IDE to collect web pages for the developer provided search queries about an exception, and then use those pages for the recommendation of relevant sections from them one by one as the developer wishes. In this way, even though Google search may return a lot of web pages, our technique can reduce the burden for the developer by recommending the relevant sections or even indicating that some particular pages might not have any relevant sections at all for the encountered exception.
We package our recommendation solution into an Eclipse plug-in prototype, called, \emph{ContentSuggest} \cite{csp}.

Our proposed technique also complements existing studies in certain aspects.
First, existing density metrics \cite{sun, maincontent} fall short in extracting content from programming related web pages, and we propose a novel density metric for programming content-- \emph{code density} 
in order to complement them.
Second, our technique introduces a novel idea of leveraging \emph{content relevance} in the extraction and then recommendation of web page content. 
It should be noted that this work is fundamentally different from our previous work--SurfClipse \cite{surfclipse} that returns a list of relevant pages for any exception. On the other hand, this work returns the most relevant sections from a given web page for the exception of interest.

\begin{lstlisting}[label=lst:ccontext, escapechar=@, language=java, belowskip=-2em, captionpos=t, aboveskip=0em,  xleftmargin=0em, firstnumber=44, frame=bt,float=t,  caption={Context code of an Exception}]
//more code goes here ...
FileInputStream fis = new FileInputStream(file);
@\hl{ObjectInputStream ois = new ObjectInputStream(fis);}@
ArrayList<Record> currentList = new ArrayList<>();
int size = ois.readInt();
for (int i = 0; i < size; i++) {
	Record current = (Record) ois.readObject();
	currentList.add(current); }
\end{lstlisting}
\begin{lstlisting}[label=lst:strace, escapechar=@, belowskip=0em, aboveskip=0em, captionpos=t, xleftmargin=0em, firstnumber=1, frame=bt,float=t, caption={Stack Trace of the Exception}]
@\hl{java.io.EOFException}@
at java.io.ObjectInputStream$PeekInputStream.readFully(ObjectInputStream.java:2325)
at java.io.ObjectInputStream$BlockDataInputStream.readShort(ObjectInputStream.java:2794)
at java.io.ObjectInputStream.readStreamHeader(ObjectInputStream.java:801)
at java.io.ObjectInputStream.<init>(ObjectInputStream.java:299)
at core.MyEOFTest.main(MyEOFTest.java:40)
\end{lstlisting}

We evaluate and validate our technique in three ways.
An experiment using 250 programming related web pages, 80 programming exceptions, and their technical details shows that
our technique recommends relevant content from a web page with a \emph{precision} of 81.96\%, a \emph{recall} of 76.74\%, and a \emph{$F_{1}$-measure} of 76.30\% on average, which are promising. We compared against 
the only available closely related technique-- \citet{sun} and found that our technique outperformed that technique in terms of all the performance metrics. 
A case study using 35 StackOverflow web pages and comparing with two state-of-the-art traceability link recovery techniques-- VSM \cite{antoniol} and LSA \cite{marcus}
reports that our technique performs significantly well in identifying the page content marked as \emph{relevant} by a large technical crowd.
Thus, the paper makes the following technical contributions:
\begin{itemize}\itemsep 0em
\item We propose a novel metric--\emph{code density} that complements existing density metrics, and extracts content from programming related web pages.
\item We introduce \emph{content relevance} in content extraction from a web page, which in turn provides a mean for supporting the developers in post-search content analysis through relevant section recommendation.
\item We package the proposed solution into an Eclipse plug-in prototype \cite{csp}, that captures the technical details of an encountered exception in the IDE, and then recommends the relevant sections from a given web page.
\end{itemize}

The rest of the paper is organized as follows-- Section \ref{sec:bg} focuses on the background concepts required for the research and Section \ref{sec:approach} discusses the proposed technique including working modules, metrics and algorithms. Section \ref{sec:experiment} describes the conducted experiments, results and validation followed by a case study, Section \ref{sec:threats} identifies the potential threats to validity, Section \ref{sec:related} focuses on the related work, and finally Section \ref{sec:conclusion} concludes the paper.

\begin{lstlisting}[label=lst:dom, escapechar=@, aboveskip=0ex, captionpos=t,  belowskip=0pt, frame=bt,float=t,  caption={Example HTML Segment (taken from \cite{domsnipp})}]
<div id="content">
<div id="question-header">
<h1 itemprop="name">
<a>How to instantiate inner class using reflection?</a>
</h1></div>
<div class="post-text" itemprop="description">
<p>I get this exception:</p>
<pre class="lang-java prettyprint prettyprinted">
<code>java.lang.InstantiationException ..</code>
</pre></div></div>
\end{lstlisting}

\begin{figure}[!tb]
\centering
\includegraphics[width=3in]{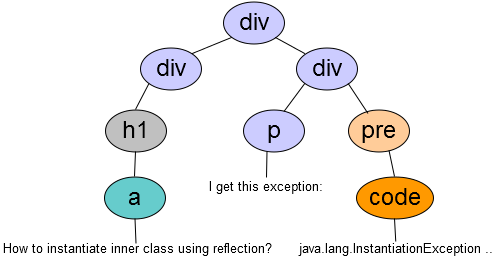}
\vspace{-.2cm}
\caption{DOM Tree of Example in Listing \ref{lst:dom}}
\label{fig:dom}
\vspace{-.5cm}
\end{figure}

\section{Background}\label{sec:bg}
\textbf{Document Object Model (DOM)}: It is a cross-platform and language independent convention to represent the content of an HTML or XML document. In this model, a document is represented as a tree, where each of the tags is represented as an \emph{inner node} and textual or graphical elements are represented as \emph{leaf nodes}. For example, the HTML code segment in Listing \ref{lst:dom} shows the title and body part of a programming question posted on StackOverflow Q \& A site, and Fig. \ref{fig:dom} shows the corresponding DOM tree. 
In our research, we use \emph{Jsoup}\footnote{http://jsoup.org}, a popular Java library, for parsing and analyzing the DOM tree of any web page.

\textbf{Cosine Similarity}:
 It is a measure that is frequently used in information retrieval in order to determine the similarity between two text documents. 
 In our research, we use cosine similarity measure for determining lexical similarity between the context (\eg\ stack trace, context code) of a programming exception and the discussion text from a candidate section of a web page. We consider each of the problem context and discussion texts as a \emph{bag of tokens}\footnote{A collection of tokens with no fixed order}, discard the insignificant tokens (\eg\ braces, semicolons, colons, dots and other punctuations), and decompose each token having a camel-case (\eg\ StringBuffer) or dotted structure (\eg\ java.io.IOException). We then prepare a combined set of tokens, $C$, from the two sets and calculate \emph{cosine similarity}, $S_{cos}$, as follows.
\begin{equation}\label{eq:cosine}
\setlength{\abovedisplayskip}{.2em}
\setlength{\belowdisplayskip}{.2em}
S_{cos}=\frac{\sum_{i=1}^{n}A_i\times B_i}{\sqrt{\sum_{i=1}^{n}A_i^2}\times \sqrt{\sum_{i=1}^{n}B_i^2}}
\end{equation}
Here, $A_i$ represents frequency of $i^{th}$ token from $C$ in set A (\ie\ exception context), and $B_i$ represents that frequency in set B (\ie\ candidate discussion text). This measure values from zero (\ie\ complete lexical dissimilarity) to one (\ie\ complete lexical similarity), and helps to determine the lexical relevance between the context of an exception and the candidate section from a web page.

\textbf{Logistic Regression}: It is a probabilistic statistical classification model that predicts binary or dichotomous outcomes based on a set of predictor variables (\ie\ features). It is widely used in medical and social science fields. In our research, we use the regression model in association with a machine learning technique for estimating the relative weights (\ie\ predictive power) of different \emph{density} and \emph{relevance} metrics for page content extraction (Section \ref{sec:weighting}). Logistic regression models the probabilities of different outcomes for a single trial as a function of predictor variables using a \emph{logistic} function. The logistic function is a common sigmoid function, $F(t)$, as follows:
\begin{equation}\label{eq:logistic}
F(t)=\frac{e^{t}}{e^{t}+1},~~t=\beta_{0}+\beta_{1}x_{1}+\beta_{2}x_{2}
\vspace{-.15cm}
\end{equation}
where $F(t)$ is a logistic function of a variable $t$, which is again a function of the predictor variables $x_{1}$ and $ x_{2}$. Here, $\beta_{1}, \beta_{2}$ are coefficients, and $\beta_{0}$ is the intercept in the regression equation. The function always returns a value between zero and one, and thus, provides a probabilistic measure for each type of the outcomes for the trial.

\begin{figure*}[!t]
\centering
\includegraphics[width=6.4in]{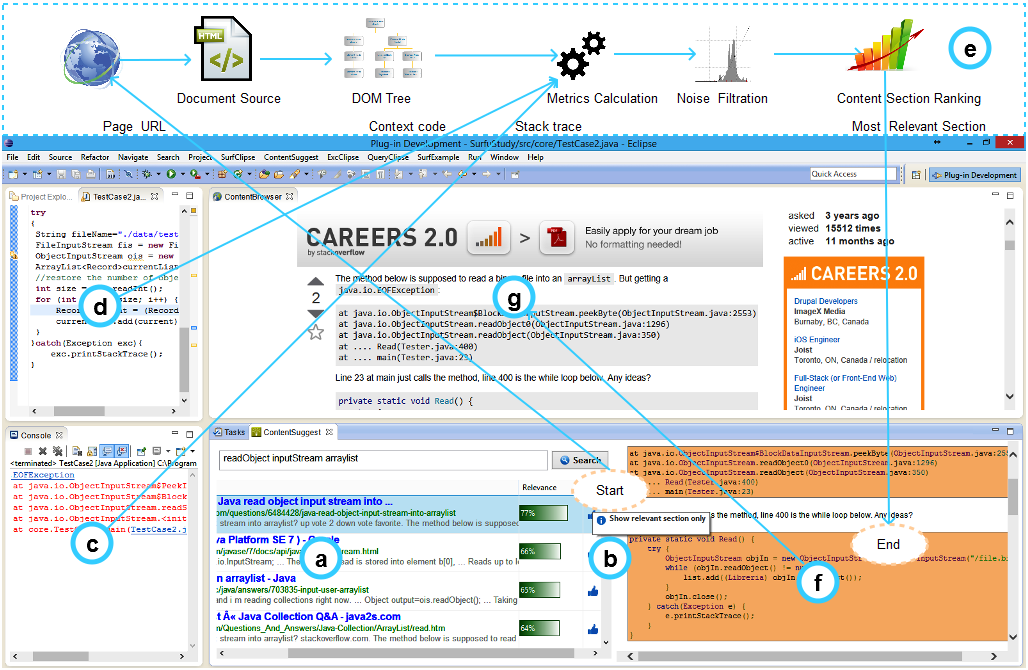}
\vspace{-.3cm}
\caption{Schematic Diagram of the Proposed Approach}
\label{fig:cssysdiag}
\vspace{-.4cm}
\end{figure*}

\section{Proposed Approach}\label{sec:approach}
\subsection{Working Modules}\label{sec:csmodule}
Our proposed technique exploits the technical details of a programming exception encountered in the IDE, and  recommends the relevant sections from a given web page. In Fig. \ref{fig:cssysdiag}, the schematic diagram of the technique shows the working modules, and explains different steps required for relevant content identification, recommendation and visualization. We package the whole solution as an Eclipse plug-in prototype \cite{csp}, and it has three modules as follows:  

{\textbf{Content Collector:} 
The \emph{collector module} collects exception message and stack trace from the active \emph{Console View} (Fig. \ref{fig:cssysdiag}-(c)) and \emph{context code}\footnote{A segment of the source code that generates the exception} from the active text editor (Fig. \ref{fig:cssysdiag}-(d)) once an exception occurs.
It also collects the HTML source of the selected web page (\eg\ top one selected in Fig. \ref{fig:cssysdiag}-(a)). Once the developer selects a web page and requests for relevant page sections, the \emph{collector module} downloads HTML source of the page, and sends the source and the previously collected exception details from the IDE to the \emph{extractor module}.

\textbf{Content Extractor:} The \emph{extractor module} (\ie\ dashed rectangle, Fig. \ref{fig:cssysdiag}-(e)) analyzes the source of the selected HTML page, parses each of the tags, and develops a DOM tree. It then analyzes each of the tree nodes, calculates their \emph{content density} and \emph{content relevance} (Section \ref{sec:metrics}), and assigns content scores. The module then discards the noisy nodes based on their content scores, and identifies the DOM tree nodes most relevant to the encountered exception (Fig. \ref{fig:cssysdiag}-(c)) in the IDE for recommendation. 

\textbf{Content Visualizer:} 
The \emph{visualizer module} consists of two content visualization panels.
First, the relevant content panel (Fig. \ref{fig:cssysdiag}-(f)) displays the most relevant section from a web page recommended by the \emph{extractor module}, where it highlights different program elements of interest such as stack trace and code segment. The idea is to help a developer instantly decide if the page is worth browsing or not. Thus, the developer can save time and effort in choosing the appropriate solution for the exception at hand. Once she is convinced by the most relevant section, she then can check the whole page using the embedded browser (Fig. \ref{fig:cssysdiag}-(g)) for in-depth analysis. 
Second, the result panel (Fig. \ref{fig:cssysdiag}-(a, b)) visualizes the estimated relevance of each result page against the target exception by analyzing the meta description of the page from the search engine. This visualization helps the developer choose the prospective solution pages in the first place during search.

\subsection{Proposed Metrics}\label{sec:metrics}
In this section, we discuss our proposed density and relevance metrics that are used for extracting and recommending relevant section(s) from  a given web page.
\subsubsection{Content Density (CTD)}
Existing studies \cite{sun, maincontent} propose two density metrics--\emph{text density}, and \emph{link density} for \emph{noise-free} content extraction from a web page. However, these metrics are based on regular texts (\eg\ news article), and they are neither properly applicable nor sufficient enough for content extraction from programming related web pages. These pages contain items other than regular texts such as stack traces, code segments, and configuration information. We thus modify existing metrics, introduce a new density metric, and then finally propose a composite density metric.

\textbf{Text Density (TD)}: \emph{Text Density} represents the amount of any textual content each of the HTML tags in the web page contains on average. 
The metric roughly estimates the content aspect of the page.
Thus, in the DOM tree, \emph{text density} ($TD_{i}$) of a node is calculated by capturing its number of child nodes ($T_{i}$) (\ie\ inner nodes) and the amount of texts ($C_{i}$) it contains in the leaf nodes as follows:   
\begin{equation}\label{eq:td}
\setlength{\abovedisplayskip}{0pt}
\setlength{\belowdisplayskip}{0pt}
TD_{i}=\frac{C_{i}}{T_{i}}
\end{equation}

\textbf{Link Density (LD)}: \emph{Link Density} represents the amount of linked (\ie\ noisy) texts each of the HTML tags contains on average. 
The metric roughly estimates the noise aspect of the page.
Existing literature \cite{sun,maincontent} considers any linked text in the web page as \emph{noise}. However, in our research, we make a careful choice about them. We analyze the relevance of each linked text element against the exception of interest, and consider the element as \emph{noise} only if its relevance is below a carefully chosen heuristic threshold ($\eta$=0.75). We otherwise consider it as a legitimate textual element. Thus in the DOM tree, the \emph{link density} ($LD_{i}$) of a node $i$ is calculated by capturing its number of child nodes ($T_{i}$) (\ie\ inner nodes) and the amount of linked or noisy texts ($LC_{i}$) it contains in the leaf nodes as follows:
\begin{equation}\label{eq:ld}
\setlength{\abovedisplayskip}{2pt}
\setlength{\belowdisplayskip}{0pt}
LD_{i}=\frac{LC_{i}}{T_{i}}
\end{equation}
We consider each \emph{<a>} tag, and check its relevance before considering it as \emph{noise}. As \citet{sun} suggest, we also consider \emph{<input>} and \emph{<button>} as linked elements, and their content as linked texts.

\textbf{Code Density (CD)}: \emph{Code Density} represents the amount of \emph{code related texts} each of the HTML tags contains on average. Programming related web pages generally contain different program elements such as \emph{stack traces} and \emph{code segments}, and they are of great interest to the developers. The developers often analyze or reuse (\ie\ code segments) them for solving their programming problems. We believe that the code related elements complement the discussion texts about programming, and thus \emph{code density} can be considered as an important indicator of legitimacy of 
a programming related web page. In the DOM tree, the  code density ($CD_{i}$) of a node $i$ is calculated by considering  its number of child nodes ($T_{i}$) (\ie\ inner nodes) and the amount of code related texts ($CC_{i}$) it contains in the leaf nodes as follows:
\begin{equation}\label{eq:cd}
\setlength{\abovedisplayskip}{1pt}
\setlength{\belowdisplayskip}{1pt}
CD_{i}=\frac{CC_{i}}{T_{i}}
\end{equation}
Previous studies \cite{surfclipse, seahawk} suggest that code related elements are generally posted in the page using \texttt{<code>, <pre>} and \texttt{<blockquote>} HTML tags. We thus consider the texts from those tags as code related texts in density calculation.

While \emph{text density} metric represents a generalized form of density for all kinds of text, both \emph{code density} and \emph{link density} point to special types of text. 
\emph{Code density} can be considered as a heuristic measure of programming elements in the text, whereas \emph{link density} is a similar measure for \emph{noise} in the content. In our research, we consider all three metrics of an HTML tag $i$, and propose a \emph{log-based composite density metric} called \emph{content density} ($CTD_{i}$). Our metric is adapted from the \emph{Composite Text Density} metric of \citet{sun}, and we choose the \emph{log-based} metric in order to better distinguish a legitimate section from a noisy section of the page. Detailed rationale of log-based density can be found elsewhere \cite{sun}.
\begin{equation}\label{eq:ctd}
\setlength{\abovedisplayskip}{0em}
\setlength{\belowdisplayskip}{1em}
\begin{split}
CTD_{i}=(TD_{i}+\frac{CD_{i}}{TD_{i}})\times \\ log_{ln(\frac{TD_{i}\times LD_{i}}{\neg LD_{i}}+\frac{LD_{b}\times TD_{i}}{TD_{b}}+e)}(\frac{TD_{i}}{LD_{i}}+\frac{CD_{i}}{TD_{i}})
\end{split}
\end{equation}
Here, $TD_{i}, CD_{i}, LD_{i}$ and $\neg LD_{i}$ represent \emph{text density}, \emph{code density}, \emph{link density} and \emph{non-link density} of the HTML tag $i$ respectively. $TD_{b}$ and $LD_{b}$ represent the \emph{text density} and \emph{link density} of \emph{body} tag respectively. In Equation \eqref{eq:ctd}, $\frac{TD_{i}}{LD_{i}}$ is a measure of the proportion of linked texts. When a tag has higher \emph{link density}, $\frac{LD_{i}}{\neg LD_{i}}\times TD_{i}$ increases the log base, $\frac{TD_{i}}{LD_{i}}$ gets a lower value, and thus overall \emph{content density} is penalized. However, $\frac{LD_{b}\times TD_{i}}{TD_{b}}$ maintains the balance between these two interacting parts, and prevents a lengthy and homogeneous text block from getting an extremely higher value or a single line text (\eg\ page title) from getting an extremely lower value. Moreover, we introduce the programming text proportion of a tag, $\frac{CD_{i}}{TD_{i}}$, which improves the overall \emph{content density} metric for the HTML tag that contains both programming texts and regular texts. 

\subsubsection{Content Relevance (CTR)}\label{sec:relevance}
Existing studies \cite{sun, maincontent} apply different density metrics in order to discard noisy sections (\eg\ advertisements) and extract  legitimate sections from a web page. However, these metrics are not sufficient enough for relevant content extraction from the web page, \ie\ our research problem. We thus leverage the technical details of an encountered exception in the IDE, and propose three relevance metrics for determining relevance of different sections from a web page.

\textbf{Text Relevance (TR)}: \emph{Text relevance} estimates relevance of the textual content from any HTML tag against a given exception and its context. The context of an exception is represented as a list of keywords collected from corresponding stack trace and context code (Section \ref{sec:contextrep}).  
For example, Listing \ref{lst:econtext} shows the context of our showcase exception--\texttt{EOFException} in Listing \ref{lst:ccontext} and Listing \ref{lst:strace}. We calculate \emph{cosine similarity} between such keyword list and the texts from each tag from the page. Cosine similarity measure represents 
the token overlap between two items. Since the context of an encountered exception contains important tokens such as class names and method names associated with the exception, lexical similarity between an HTML tag and the context suggests the tag's relevance for the exception.
The similarity measure values from zero to one, where one refers to complete lexical relevance and vice versa.


\textbf{Code Relevance (CR)}: \emph{Code relevance} estimates relevance of a code segment or a stack trace block from an HTML tag against corresponding context code or stack trace of a given exception. 
In order to estimate \emph{code relevance} of a node from the DOM tree,
we analyze three types of child tags-- \texttt{<code>},\texttt{<pre>} and \texttt{<blockquote>} under that node. According to traditional heuristics \cite{surfclipse}, such tags generally contain the program elements (\eg\ code segments).
We apply two different techniques for stack traces and code segments for estimating their relevance with their counterparts.

Stack trace of a programming exception contains an error message followed by a list of method call references that point to the possible error locations in the code. 
We develop separate token list by collecting suitable tokens (\eg\ class name, method name) from each of the stack trace blocks of an HTML tag and the stack trace in the IDE respectively. We then calculate \emph{cosine similarity} between the two token lists, and consider the measure as an estimate of relevance 
for the tag to the exception in the IDE.


In order to estimate relevance of a code segment from the HTML tag against an exception of interest, we collect the \emph{context code} of the exception, and apply a state-of-the-art code clone detection technique by \citet{croy}. The technique finds out the \emph{longest common subsequence} of source tokens ($S_{lcs}$) between two code segments. We then use it for determining similarity of the code segment from the HTML tag with the context code as follows, where $S_{total}$ refers to the sequence of all tokens collected from the context code of the target exception.
\begin{equation}\label{eq:ccx}
S_{ccx}=\frac{|S_{lcs}|}{|S_{total}|}
\end{equation}

Once the relevance of all the program elements-- stack traces and code segments under an HTML tag are estimated, we find the  maximum estimate, and consider it as the \emph{code relevance} for the tag.
The metric helps in separating a highly relevant HTML tag containing relevant code elements from a less relevant tag in the web page.

While \emph{text relevance} focuses on the relevance of any textual element within an HTML tag, \emph{code relevance} estimates the relevance of program elements within it. We combine both relevance metrics in order to determine the \emph{composite relevance metric} called \emph{content relevance} ($CTR$) as follows:
 \begin{equation}\label{eq:ctr}
CTR_{i}=\alpha \times TR_{i}+\beta \times CR_{i}
\end{equation}
Here $\alpha$ and $\beta$ are the relative weights (\ie\ predictive power) of the corresponding relevance metrics. We consider a heuristic value of 1.00 for $\alpha$ and 0.59 for $\beta$, and they
are estimated using a machine learning based technique (Section \ref{sec:weighting}).
\begin{lstlisting}[label=lst:econtext,  belowskip=-12pt, aboveskip=1em,  xleftmargin=0em, firstnumber=44, frame=bt,float=t,  caption={Context of the Exception in Listing \ref{lst:ccontext} and Listing \ref{lst:strace}}]
Exception in thread "main" java.io.EOFException readInt  
ObjectInputStream readStreamHeader BlockDataInputStream  
readObject readShort add main readFully FileInputStream Record ArrayList PeekInputStream init
\end{lstlisting}

\subsection{Content Score (CTS)}\label{sec:score}
We consider two different aspects--\emph{density} and \emph{relevance} for each of the content sections in the page for extracting the relevant ones.
While the density metrics focus on the legitimacy (\ie\ purity) of the content in the page, 
relevance metrics check the relevance of the same content section against the programming problem (\ie\ encountered exception) at hand.
The idea is to recommend such section of a page to the developers that is both legitimate (\ie\ noise-free) and relevant (\ie\ discusses similar problem).
We thus combine both aspects, normalize corresponding metrics from Section \ref{sec:metrics}, and propose a composite score metric called \emph{content score} ($CTS_{i}$) for each of the tags from the page as follows:
 \begin{equation}\label{eq:cs}
CTS_{i}=\gamma \times CTD_{i}+\delta \times CTR_{i}
\end{equation}
Here $\gamma$ and $\delta$ are  relative weights (\ie\ predictive power) of the corresponding density and relevance metrics--\emph{content density (CTD)} and \emph{content relevance (CTR)}.
In our experiments, we note that our technique performs the best when the equal weight ($\gamma=\delta=1.00$) is assigned to both metrics. 
The weight estimation process can be found in Section \ref{sec:weighting}.
\subsection{Extraction of Relevant Page Section(s)}\label{sec:cntext}
An HTML page is generally divided into a set of identifiable sections (\ie\ tags) that can be represented as the child nodes under \texttt{body} node in the corresponding DOM tree. 
Our contribution lies in identifying the most relevant section(s) from that page.
Once content score (Section \ref{sec:score}) for each of the tags (\ie\ nodes) in the page (\ie\ DOM tree) is calculated, we filter the tree nodes using a heuristic threshold. We consider the content score of \texttt{body} node as the threshold score, as suggested by \citet{sun} for density-based extraction. We preserve the child nodes under \texttt{body} node in the tree having scores greater than the threshold while discarding the others. We then explore each of the preserved child nodes, and find out the inner node with the highest content score. The highest score of the node indicates that the corresponding tag in the HTML page contains the most salient content in terms of legitimacy and relevance for the programming problem at hand. In order to discard noisy or less important elements, we keep that highest scored node along with its child nodes, and mark them as \emph{content} whereas the remaining siblings are marked as \emph{noise}.
We apply the same process recursively for each node in the DOM tree, and finally we get each node in the tree annotated as either \emph{content} or \emph{noise}. Then we discard the noisy nodes, and extract the HTML tags corresponding to the remaining nodes in the tree as the \emph{noise-free} sections of the page \cite{sun}.

Since the first step extracts several sections that might not be equally relevant, we need further filtration for collecting the highly relevant section(s) from the page.
We thus focus on \emph{content relevance} (Section \ref{sec:relevance}) of the preserved nodes, and choose the node with the highest content relevance for recommendation. This highest relevance score indicates that the corresponding HTML tag is relevant both in terms of programming content and discussion texts. 
We thus ensure that the recommended sections from the page are not only relevant to the problem at hand (\ie\ exception) but also are legitimate enough in content to survive the noise filtration.

For example, our proposed technique returns these metric values-- \emph{TD}=32.74, \emph{LD}=2.88, \emph{CD}=24.29, \emph{CTD}=0.02, \emph{CR}=0.84, \emph{TR}=0.83, \emph{CTR}=0.99, and \emph{CTS}=1.0144, for the page section in Fig. \ref{fig:relsec}. This section outperforms other sections in Fig. \ref{fig:moti} both in terms of legitimacy (\ie\ purity) and relevance with the target exception in Listing \ref{lst:strace}.
Thus, our technique marks the section as the highly relevant one, and extracts it for recommendation.

\subsection{Metric Weight Estimation}\label{sec:weighting}
In order to determine relative weights of two relevance metrics--\emph{text relevance} and \emph{code relevance}  and two composite metrics--\emph{content density} and \emph{content relevance}, we choose 50 random web pages from the dataset.
Details on dataset preparation can be found in Section \ref{sec:dataset}. We then collect the corresponding metrics for 33,360 text blocks (\ie\ tags) from those pages using our technique. We identify whether each of those blocks is included in the \emph{gold content} or not (Section \ref{sec:dataset}), which provides a \emph{binary class label} (\ie\ "0" or "1") for the text block against its set of metrics (\ie\ features). 
We then apply machine learning on the collected block samples using logistic regression that provides a regression model \cite{specmining}
The model is developed on Weka\footnote{http://www.cs.waikato.ac.nz/ml/weka}, and it is validated using 10-fold cross-validation.
The regression model contains a coefficient for each of the features which are tuned by Weka for classifying each sample with maximum accuracy. 
We believe that these coefficients are an estimate of the predictive power for the features used in the model, and we consider them as the weights of the individual relevance metrics \cite{specmining}. 
For the sake of simplicity and for reducing bias, we normalize those coefficients, and consider a heuristic weight $\alpha$=1.00 for \emph{text relevance} and $\beta$=0.59 for \emph{code relevance} metrics. While $\beta$ has an initial value of 0.86 from the regression model, we got the global maximum at $\beta=0.59$ for our dataset through iterative experiments \cite{surfclipse}. 

In case of composite density and composite relevance metrics, we find that the proposed technique performs significantly well with equal relative weights assigned. Thus, we consider a heuristic weight of 1.00 for both of the composite metrics, \ie\ $\gamma=\delta=1.00$.

\subsection{Exception Context Representation}\label{sec:contextrep}
In our research, we not only take the density metrics but also the relevance estimate of each of the sections from the web page into consideration. 
Each page in the dataset (Section \ref{sec:dataset}) is relevant to a particular exception, and we exploit the details such as stack trace and context code of that exception in the dataset for relevance estimation of different sections from the page. We analyze the stack trace (\eg\ Listing \ref{lst:strace}) and extract different tokens such as \emph{package name}, \emph{class name} and \emph{method name} from each of the method call references. We also analyze the context code (\eg\ Listing \ref{lst:ccontext}) of the exception and collect \emph{class} and \emph{method name} tokens. We use \emph{Javaparser}\footnote{http://code.google.com/p/javaparser/} for compilable code and an \emph{island parser} for non-compilable code for extracting the tokens \cite{surfclipse}. Then we combine tokens from both the stack trace and the context code, and append the exception name along with the exception message (\ie\ highlighted line of the stack trace in Listing \ref{lst:strace}) to the combined set. We call this token set as the \emph{context representation} for the exception of interest. For example, Listing \ref{lst:econtext} shows the context representation by our technique for an \emph{EOFException} with stack trace in Listing \ref{lst:strace} and context code in Listing \ref{lst:ccontext}. We use such context representation to estimate the relevance of any page sections to the exception (Section \ref{sec:relevance}).

\section{Evaluation \& Validation}\label{sec:experiment}

\subsection{Experimental Dataset}\label{sec:dataset}
\textbf{Data Collection:} We use a dataset of 250 web pages and 80 programming exceptions associated with standard Java platform and Eclipse plug-in framework for experiments.
We include the technical details of each exception such as stack trace and context code in the dataset.
For details on how exceptions  were collected, please consult our previous work \cite{surfclipse}.  
It should be noted that each of the pages is carefully selected and taken from the dataset of the previous study \cite{surfclipse}. 
We also collect the HTML source of each of the pages, and include in the dataset for the experiments.

\textbf{Gold Set Development:} We manually analyze each of the 250 pages, and extract \emph{gold content} from them for the study.
We consider the most relevant page section for a given exception as the gold content from the page.
We also adopt a simplified definition for relevant sections in the page.
Programming sites focusing on errors and exceptions often include code snippets and stack traces as a part of discussion.
We look for such page sections that contain relevant stack traces or relevant code segments, and extract them as the gold content through an extensive manual analysis of 20 to 25 working hours. 
One can wonder about the simplified definition of relevance given that the concept of relevance is mostly subjective.
However, our goal in this work is to present the presumably relevant sections to signal the relevance of a selected page, and help the developers find the solution with less information analyzed.
Thus, the adoption of simplified relevance for page sections is justified.

\textbf{Cross-Validation:} Since the concept of relevance is subjective, in order to reduce the bias, we perform cross validation on the gold set with the help of peers.
Two graduate research students randomly checked a subset of ten pages each, and submitted the most relevant sections from the pages against the exceptions of interest.
We found that most of their choices match with our gold set selection which provides us confidence on the data.
We made this gold set available online \cite{csp} for others to use.

\textbf{Data from Stack Overflow:} About 40\% of the pages came from StackOverflow (SO) site, and we thus divide the pages into two subsets called \emph{SO-Pages} and \emph{Non-SO Pages}. Again all data are hosted online \cite{csp}.
While both sets are used for evaluation and validation, we additionally use \emph{SO-Pages} for a case study (Section \ref{sec:casestudy}) where our technique locates the \emph{highest voted} and the \emph{accepted} answer posts from a given StackOverflow page.

\subsection{Performance Metrics}\label{sec:perm}
Our proposed technique is greatly aligned with the research areas of information retrieval and recommendation systems, and we use a list of performance metrics from those areas for evaluating our technique as follows \cite{sun, surfclipse}:

\textbf{Mean Precision (MP)}: \emph{Precision} determines the percentage of the retrieved content that is expected (\ie\ in the gold content) from a web page. In our evaluation, we compare the retrieved content by our technique with the manually extracted gold content. As \citet{sun} suggest, we use longest common subsequence (LCS) of words between retrieved content and gold content for precision calculation. Thus, \emph{precision} can be determined as follows, where \emph{a} refers to the word sequence of retrieved content and \emph{b} refers to that of the corresponding gold content. 
\begin{equation}\label{eq:mp}
\setlength{\abovedisplayskip}{0em}
\setlength{\belowdisplayskip}{0em}
P=\frac{|LCS(a, b)|}{|a|},~~ MP=\frac{\sum_{i=1}^{N}P_{i}}{N}
\end{equation}
\emph{Mean Precision (MP)} averages the precision measures for all web pages ($N$) in the dataset.

\textbf{Mean Recall (MR)}: \emph{Recall} determines the percentage of the expected content  (\ie\ gold content) that is retrieved from a web page by a technique. We calculate the \emph{recall} of a technique as follows:
\begin{equation}\label{eq:mr}
\setlength{\abovedisplayskip}{0em}
\setlength{\belowdisplayskip}{1pt}
R=\frac{|LCS(a, b)|}{|b|}, ~~ MR=\frac{\sum_{i=1}^{N}R_{i}}{N}
\end{equation}
\emph{Mean Recall (MR)} averages the \emph{recall} measures for all pages ($N$) in the dataset.

\textbf{ Mean $F_{1}$-measure (MF)}: While each of \emph{precision} and \emph{recall} focuses on a particular aspect of the performance of a technique, $F_{1}$-measure is a combined and more meaningful metric for evaluation\footnote{http://stats.stackexchange.com/questions/49226/}. We calculate $F_{1}$-measure from the harmonic mean of \emph{precision} and \emph{recall} \cite{sun} as follows: 
\begin{equation}\label{eq:mf}
\setlength{\abovedisplayskip}{0em}
\setlength{\belowdisplayskip}{1pt}
F_{1}=\frac{2 \times P \times R}{P+ R},~~MF=\frac{\sum_{i=1}^{N}F_{1i}}{N}
\end{equation}
\emph{Mean $F_{1}$(MF)} averages all such measures.


\subsection{Experimental Results} \label{sec:results}
We conduct experiments on the proposed technique using our dataset, and evaluate the technique using three performance metrics--\emph{precision}, \emph{recall} and \emph{$F_1$-measure}. 
Our technique extracts relevant content from the web pages with a \emph{mean precision} of 81.96\%, a \emph{mean recall} of 76.74\%, and a \emph{mean $F_{1}$-measure} of 76.30\%. 
Table \ref{table:srcomp} investigates the effectiveness of the two aspects--\emph{density} and \emph{relevance} associated with the page content in extracting relevant sections.
We consider each of these aspects in isolation as well as in combination, and evaluate our technique with different sets of web pages.
In case of \emph{content density}, the proposed technique performs well in terms of \emph{recall} and performs significantly poor in terms of \emph{precision} with all three sets--\emph{StackOverflow pages, Non-StackOverflow pages} and \emph{All pages}. For example, the technique can return only 50.07\% relevant content (\ie\ \emph{precision}) while it uses \emph{density} metrics alone. In the case of \emph{content relevance} metric, our technique extracts relevant content from a web page with relatively better \emph{precision} (\eg\ 76.23\%), but the \emph{recall} rates are still poor (\eg\ 55.44\%). 
On the other hand, when we combine both the density and relevance metrics, we experience significant improvements in all three performance metrics with each of the sets of web pages. 
For example, our technique successfully extracts 76.74\% of the gold content with 81.96\% precision when both metrics are considered in combination. This clearly shows the benefit of our introduced paradigm--\emph{content relevance} in the extraction of relevant content from a web page, which is one of our primary objectives of this work. The finding also justifies our use of various density and relevance metrics since it demonstrates their isolated and combined effectiveness in relevant content extraction.

Among the two subsets, our technique performs comparatively better for \emph{StackOverflow pages} with the metrics both in isolation and in combination. During gold set development, we note that StackOverflow pages are  structurally organized in the presentation of questions and answers, and they contain relatively less noise. This might have helped our technique perform better. However, we do further investigation with SO-Pages in Section \ref{sec:casestudy}.



\begin{table}[!t]
\caption{Experimental Results for Different Metrics}
\label{table:srcomp}
\centering
\resizebox{3.3in}{!}{%
\begin{threeparttable}
\begin{tabular}{l|l|c|c|c}
\hline
\textbf{Score Combination} & \textbf{Metric} & \textbf{SO Pages} & \textbf{Non-SO Pages} & \textbf{All Pages}\\
\hline
\multirow{3}{*}{\{Content Density (CTD)\}} & MP& 50.91\%& 49.50\% & 50.07\%\\
\hhline{~----}
 & MR& 91.74\% & 75.71\% &\textbf{82.18}\%\\
\hhline{~----}
 & MF& 62.32\% & 53.76\% & 57.22\%\\
\hhline{-----}
 \multirow{3}{*}{\{Content Relevance (CTR)\}} & MP & 86.63\% & 69.17\% & \textbf{76.23}\%\\
\hhline{~----}
 & MR& 52.17\% & 57.66\% &55.44\%\\
\hhline{~----}
 & MF& 61.07\% & 55.88\%& 57.98\%\\
\hhline{-----}
 \multirow{2}{*}{\{Content Density (CTD) \&} & MP &  92.64\% & 74.60\% & \textbf{81.96}\%\\
\hhline{~----}
 & MR& 74.17\% & 78.51\% & \textbf{76.74}\%\\
\hhline{~----}
 Content Relevance (CTR)\} & MF& 80.95\% & 73.09\% & \textbf{76.30}\% \\
\hline
\end{tabular}
\center
\textbf{MP}=Mean Precision, \textbf{MR}=Mean Recall, \textbf{MF}=Mean $F_1$-measure
\end{threeparttable}
}
\vspace{-.5cm}
\end{table}

\subsection{Comparison with Existing Approaches}\label{sec:compare}
Since there is no existing study that addresses the same research problem as ours, \ie\ relevant section(s) extraction from programming related web pages, we choose a closely related existing technique-- \citet{sun}. It applies a list of density metrics for extracting \emph{noise-free} content from a given web page. We replicate their technique with minor adjustments in our working environment, and collect the most legitimate (\ie\ representative) section of the page extracted by the technique.
The idea is to investigate how closely the legitimate section by \citet{sun} matches with the manually extracted relevant section (\ie\ gold content), and also to validate the performance of our technique.
We compare with their technique for the same dataset, and find out that our technique performs comparatively better in terms of all performance metrics. 
Table \ref{table:compare} and Fig. \ref{fig:compare} report our findings from the comparative study.

Fig. \ref{fig:compare} shows the comparative analysis between the two techniques using box plots. We note that our technique performs significantly better in terms of especially \emph{precision} and \emph{f-measure} than \citet{sun}. Our technique provides a \emph{median} measure from 85\% to 100\% for all three metrics, whereas their technique provides such measure from 40\% to 50\% with an exception in \emph{recall} that ranges around 80\%. Table \ref{table:compare} further breaks down the results into different subsets-- \emph{SO-Pages} and \emph{Non-SO Pages}, and we experience the similar findings. While all these findings demonstrate the potential of our technique in locating relevant sections in the page, 
one could still argue about the relevance of the extracted section, in particular, because of the subjectivity involved.
This concern is addressed using a case study in Section \ref{sec:casestudy}.

\begin{figure}[!t]
\centering
\includegraphics[width=3.2in]{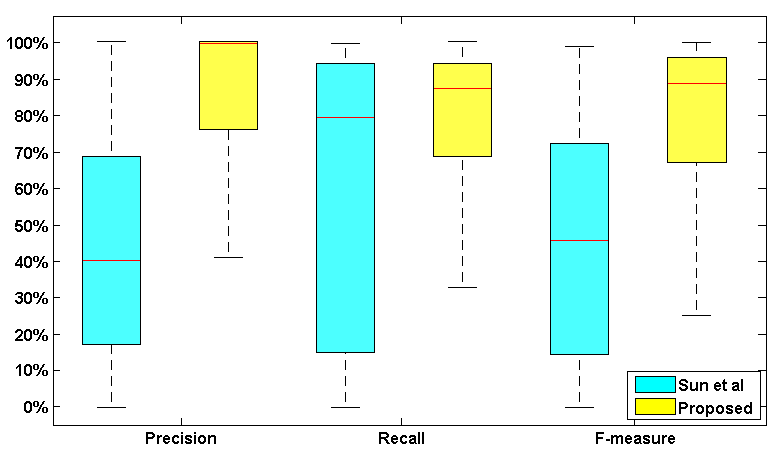}
\vspace{-.3cm}
\caption{Comparison with Existing Technique}
\label{fig:compare}
\vspace{-.3cm}
\end{figure}

\begin{table}[!t]
\caption{Comparison with an Existing Technique}
\label{table:compare}
\centering
\resizebox{3.3in}{!}{%
\begin{threeparttable}[b]
\begin{tabular}{l|l|c|c|c}
\hline
\textbf{Content Extractor} &\textbf{Metric} & \textbf{SO Pages} & \textbf{Non-SO Pages} & \textbf{All Pages (D)}\\
\hline
\multirow{2}{*}{\citet{sun}} &MP & 52.63\% & 38.89\% & \textbf{44.44}\%  \\
\hhline{~----}
 & MR& 86.49\% & 41.84\% & \textbf{59.88}\%  \\
\hhline{~----}
(Density) & MF & 62.57\%& 34.49\% & \textbf{45.84}\% \\
\hhline{-----}
 \multirow{2}{*}{Proposed Approach} & MP &  92.64\% & 74.60\% & \textbf{81.96}\%\\
\hhline{~----}
 & MR& 74.17\% & 78.51\% & \textbf{76.74}\%\\
\hhline{~----}
 \{Density \& Relevance\}& MF& 80.95\% & 73.09\% & \textbf{76.30}\% \\
\hline
\end{tabular}
\end{threeparttable}
}
\vspace{-.4cm}
\end{table}

\subsection{Case Study with StackOverflow Pages}\label{sec:casestudy}
Although our gold set for the experiments is carefully prepared and validated by the peers, it may still contain subjective bias. Thus, the evaluation and the validation results might be biased. In order to mitigate this threat, we exploit an alternative approach, and conduct a case study using StackOverflow questions and answers. This section discusses the details of the conducted case study.

\textbf{Dataset Preparation:}
StackOverflow API\footnote{http://data.stackexchange.com/stackoverflow} provides access to a rich dataset of questions and answers, and we exploit that API service in developing our dataset for the case study. The goal is to investigate whether our proposed technique can actually locate the answer posts within the page that are either accepted as solutions or highly voted by the large user base of StackOverflow. In StackOverflow, each of the posted answers is  reviewed by thousands of technical users, and we leverage that crowd knowledge (\ie\ the evaluation by hundreds, if not thousands of users) in developing the gold set for this case study. We chose 35 StackOverflow questions related to 29 programming exceptions from our dataset based on this condition-- each of the questions should have at least three answers with one answer accepted as \emph{solution}. We develop two gold sets-- \emph{most-voted-gold-set} (\ie\ contains top-scored answers) and \emph{accepted-gold-set} (\ie\ contains accepted answers) for the study. While the question section is discarded from the StackOverflow page for reducing noise, we preserve the DOM structure of the page for enabling our technique to operate conveniently and to leverage the structure for relevant section extraction.

\textbf{Running of Study:}
One can essentially think of our research problem as a standard traceability problem since our technique basically links the encountered exception (and its details) to the relevant sections in the web page. 
However, there exist several important particulars which need to be considered. First, our technique attempts to identify the most relevant section(s) from within a single HTML page rather than an entire page (\ie\ document) from within a large corpus.
Second, automatically separating relevant sections from an HTML page is a non-trivial task, and our technique exploits the DOM structure of the page for extracting such sections.
We apply our technique on each of 35 StackOverflow pages, extract the most relevant sections, and then compare them with the most voted and accepted answer posts from the gold set.
Two state-of-the-art information retrieval techniques-- Latent Semantic Analysis (LSA) and Vector Space Model (VSM) are successfully applied in traceability link recovery by several existing studies \cite{antoniol, marcus,bavota}, and 
we also contrast our technique with them.
Since those techniques require a corpus for information retrieval, 
we represent each of the answer posts from the page as an individual document in the corpus for that page. We then use the \emph{context} (Section \ref{sec:contextrep}) of the exception related to that page as the search query for retrieving the most relevant document (\ie\ answer post).
For {LSA}, we use \emph{TML}\footnote{http://tml-java.sourceforge.net}, a text mining library, and for VSM, we use \emph{Apache Lucene}\footnote{http://lucene.apache.org/core}, a popular VSM-based search engine.

\begin{table}[!t]
\caption{Comparison with Existing IR Techniques}
\label{table:ircompare}
\centering
\resizebox{3.3in}{!}{%
\begin{threeparttable}[b]
\begin{tabular}{l|l|c|c}
\hline
\textbf{Content Extractor} &\textbf{Metric}  &\textbf{Accepted Posts} & \textbf{Most Voted Posts}\\
\hline
\multirow{3}{*}{Latent Sematic Analysis \cite{marcus}} &MP  & 19.98\% & \textbf{23.02}\% \\
\hhline{~---}
 & MR & 21.78\% & \textbf{23.17}\%\\
\hhline{~---}
 & MF & 18.43\% & 21.07\%\\
\hhline{----}
\multirow{3}{*}{Vector Space Model \cite{antoniol}} &MP  & 22.50\%  & \textbf{33.89}\%\\
\hhline{~---}
 & MR & 23.08\% & \textbf{31.90}\%\\
\hhline{~---}
 & MF & 19.77\% & 30.44\%\\
\hhline{----}
\multirow{2}{*}{Proposed Approach} & MP  & 23.10\% & \textbf{31.36}\%\\
\hhline{~---}
 & MR & 45.15\% & \textbf{54.42}\%\\
\hhline{~---}
 \{Density \& Relevance\}& MF & 26.99\% & \textbf{35.90}\%\\
\hline
\end{tabular}
\center
\textbf{MP}=Mean Precision, \textbf{MR}=Mean Recall, \textbf{MF}=Mean $F_1$-measure
\end{threeparttable}
}
\vspace{-.4cm}
\end{table}

\begin{figure}[!t]
\centering
\includegraphics[width=3.2in]{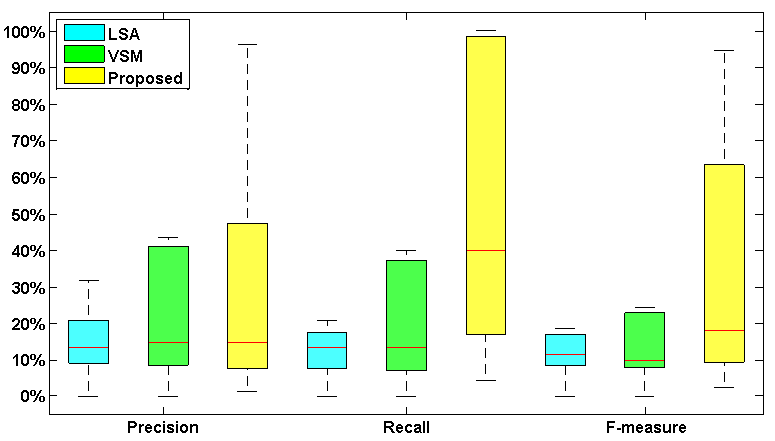}
\vspace{-.3cm}
\caption{Comparison with existing IR techniques}
\label{fig:ircompare}
\vspace{-.5cm}
\end{figure}

\textbf{Results and Discussions:}
From Table \ref{table:ircompare}, we note that our technique performs comparably in terms of \emph{precision} and significantly better in terms of \emph{recall} than the other two techniques. For example, our technique returns 54.42\% of the gold content with a \emph{precision} of 31.36\%, which is promising. One can argue that the results are relatively poor compared to our previously reported results (Section \ref{sec:results}), which is true. However, we would argue that these results are still promising according to the relevant literature \cite{antoniol, marcus}, and our technique actually performs better than the two state-of-the-art information retrieval techniques-- LSA and VSM. They can retrieve at most  23.17\% and 31.90\% of the gold content (\ie\ \emph{most-voted-gold-set}) respectively. 
On the other hand, our technique is found more effective in automatically locating the answer post within a given page which is reported as the most helpful (\ie\ most up-voted post) by thousands of technical users from StackOverflow.
As shown in Table \ref{table:ircompare}, our technique also locates the answer post accepted as solution by the users from a given SO page more effectively than the other two competing techniques.

Fig. \ref{fig:ircompare} summarizes the comparative analysis among the three techniques using box plots. We note that our technique is comparable to \emph{LSA} and \emph{VSM} in terms of \emph{precision} and  significantly better in terms of \emph{recall}.
The \emph{median recall} of our technique ranges from 40\% to 50\% whereas its counterparts in the remaining techniques range from 10\% to 20\%. Since \emph{F-measure} combines both \emph{precision} and \emph{recall}, we report that our technique actually performs better as a whole due to its improved \emph{recall} rates. It also should be noted that our technique automatically locates the gold answer sections by exploiting the DOM structure of the whole page. On the other hand, those techniques operate on already extracted answer sections from the Stack Overflow page, which provides their comparable precision. Thus, despite of smaller sample size and relatively lower performance than that of first experiment (Section \ref{sec:results}),
the finding clearly demonstrates the potential of our technique for relevant and useful content recommendation.

\section{Threats to Validity}\label{sec:threats}
In our research, we note a few issues worthy of discussion. 
First, the lack of a fully-fledged user study for evaluating usability of the technique is a potential threat. 
However, our objective was to focus on the technical aspects of the approach. Furthermore, in order to at least partially evaluate the usability, 
we conduct a limited user study with five participants, where three of them have professional software development experience.
We ask them six questions about our relevance visualization, highlighting of the artifacts of interest, and IDE-based information search. 
Five out of five participants responded and suggested that the proposed technique is likely to be really helpful in extracting the desired information from a web page. However, a fully-fledged user study is required to explore the actual usability of our technique that we consider as a scope of future study.

Second, the dataset (Section \ref{sec:dataset}) prepared for evaluation and validation may contain subjective bias. In order to reduce the bias, we perform cross-validation with the help of peers, analyze their suggestions and then finalize the gold set. 
More importantly, we conduct a case study with StackOverflow pages, where gold sets are prepared by exploiting the feedback from thousands of technical users.  The study also demonstrates the potential of our technique against two traceability link recovery techniques-- LSA and VSM.

Third, metric weights are estimated using a limited training dataset (Section \ref{sec:weighting}) that might cause weight overfitting. However, we also tune and test the weights significantly against different set of pages
to mitigate the threat.

\section{Related Work}\label{sec:related}
A number of existing studies are conducted on web page content extraction, and they apply different techniques such as template or similar structure detection \cite{boilerplate, cafella}, machine learning \cite{mladv, mlnepo, densitometric}, information retrieval, domain modeling \cite{little}, and  page segmentation and filtration \cite{maincontent, sun, densitometric, vips, dsc, ccb}. The last group of techniques using page segmentation and noise filtration are closely related to our research in terms of working methodologies although they are driven by different goals. In order to extract \emph{noise-free} content from a web page, they apply several density metrics and link element based heuristics. On the other hand, we complement those density metrics, introduce novel relevance metrics, and then combine both metrics for relevant section extraction from a web page.
In particular, we recommend relevant sections from the page for an encountered exception in the IDE.
\citet{sun} exploit link elements (\eg\ \texttt{<a>, <input>} tags) for the filtration of noisy sections in a web page. This is probably ideal for news-based websites. However, the idea may not be properly applicable for programming related web sites as our experimental results suggest (Section \ref{sec:compare}). 
The technique by \citet{dsc} is actually designed with a table-based architecture of the web page in mind, which may not be applicable for modern complex websites. The two versions of \emph{Code Content Blurring} by \citet{ccb} are only tested against the news-based websites containing simple structures and homogeneous texts.

The other studies use different methodologies that are not closely related to our work, and we do not compare against them in our experiments. \citet{little} analyze real estate websites and extract property or price related information. They exploit a domain-specific model for content extraction which might not be applicable for programming related websites.
\citet{densitometric} analyze news-based websites, extract different densitometric features, and apply a machine learning classifier (C4.5) for classifying the legitimate and noisy content sections. Their approach is subject to the amount and quality of training data as well as the performance of the classifier. 
\citet{cafella} focuses on Wikipedia pages, identifies the special structures (\eg\ tabular), and mines different factual information (\eg\ list of American presidents) from the pages.
Thus, while other techniques focus on extracting the noise-free sections or mining the factual or commercial data from news, real estate or Wikipedia pages, our technique attempts to support software developers in collecting relevant information for problem at hand from the programming related web pages. 

This work is to some extent similar to our previous work--SurfClipse \cite{surfclipse} since both of them analyze exceptions and web pages. 
However, this work--ContentSuggest is also significantly different from our previous work that returns a list of relevant pages for any exception. On the other hand, this work returns the most relevant section(s) from a given web page for the exception of one's interest. 
From technical point of view, it proposes a novel metric--\emph{content relevance} for relevant section extraction, which was not considered by any of the existing approaches.
Our technique not only extracts the noise-free sections but also directs a developer to the right (or relevant) sections in the page by exploiting the details of an exception encountered in the IDE, which is a novel idea of developer support, and is not provided by any of the existing approaches.

There exist also several studies \cite{antoniol, marcus, bavota} in the literature that use information retrieval techniques for traceability link recovery, and they are also related to our work to some extent. While most of them focus on establishing links from software artifacts such as source code to software documentations or requirement documents, we attempt to link an encountered exception (and its details) to the most relevant or the most helpful section from a given web page. For detailed comparison with information retrieval techniques, we refer the readers to Section \ref{sec:casestudy}.

\section{Conclusion}\label{sec:conclusion}
To summarize, we propose a novel recommendation technique that recommends the most relevant section(s) from a given web page
for an encountered exception in the IDE.
Experiments with 250 web pages and 80 programming exceptions show that 
our technique can extract relevant content with a precision of 81.96\%, a recall of 76.74\% and a $F_1$-measure of 76.30\%, which are promising. Comparison with the only available closely related technique also shows that our technique performs significantly better in all performance metrics.
Finally, a case study with StackOverflow pages, where we compare with two state-of-the-art traceability link recovery techniques, shows that our technique is highly promising in identifying the top-scored answer posts from a StackOveflow Q \& A page by using the proposed metrics as well.
\balance
\bibliographystyle{plainnat}
\scriptsize
\setlength{\bibsep}{0.0pt}
\bibliography{sigproc}  

\end{document}